\documentclass[amssymb,twoside,12 pt]{article}
\usepackage{latexsym,amsmath}
\usepackage{hyperref}
\usepackage{amsfonts}
\usepackage{color}
\usepackage{natbib}
\usepackage{amsmath}
\usepackage{dsfont}
\usepackage{enumerate}

\usepackage{amssymb}
\usepackage[center]{titlesec}
\usepackage{graphics}
\title{ Vacuum Solutions of FRW and Axially Symmetric space-time in $f(R)$ Theory of Gravity}

\author{P. K. Agrawal, D. D. Pawar \\
School of Mathematical Sciences,\\
Swami Ramanand Teerth Marathwada University, Nanded-431606, India\\
E-mail: agrawalpoonam299@gmail.com, dnyaneshwarpawar@srtmun.ac.in }
\date{\today}

\begin{document}
\setcounter{page}{1}
\maketitle
\begin{abstract}
In the present paper an attempt has been made to study the spatially homogeneous and isotropic FRW model and axially symmetric space-time in $f(R)$ theory of gravity. We have obtained the solutions of the field equations in vacuum. To find the solutions of axially symmetric space-time we have assumed the relation between scale factor and scalar field and also we considered that scalar expansion $(\theta)$ in the model is proportional to shear scalar $(\sigma)$. The physical behavior for both models have been discussed by using physical parameters. The function of Ricci scalar is investigated for these models. \\\\
\noindent \textbf{Key words:} $f(R)$ theory of gravity, FRW universe, Axially symmetric space-time. \\
\end{abstract}

\def\baselinestretch{1.5}
\allowdisplaybreaks
\section{Introduction}
Nowadays the accelerating expansion of the universe has attracted much attention. Cosmologist have investigated that dark energy is the most important factor behind this expansion of the universe, with the help of some observational data like Type Ia Supernovae (SNeIa 8)  \citep{1538-3881-116-3-1009,0004-637X-517-2-565}, the Cosmic Microwave Background (CMB) \citep{0067-0049-148-1-1,PhysRevD.69.103501}, Large Scale Structure, Observational Hubble constant data etc. In physical cosmology, dark energy (DE) is an unknown form of energy. 
Though not very dense, it is assumed that DE fills the otherwise empty space uniformly constituting almost 68\% of universal density. Even though the physical nature of DE is still unknown, it is the most accepted hypothesis to explain the observations indicating accelerated expansion of the universe. The DE can have three forms namely, cosmological constant (vacuum energy) model, quintessence model and phantom model which are time dependent and depends on the values of the EoS parameter $\omega$ (see e.g. \cite{2004LNP...653..141S} and the references therein).
\paragraph{}To examine the nature of DE several modified theories of gravity like $f(R)$ gravity, $f(G)$ gravity, $f(T)$ gravity, $f(R,T)$ gravity have been investigated. Among these theories $f(R)$ gravity is an extension of Einstein's General Relativity. Actually $f(R)$ gravity is a family of theories in which different function of Ricci scalar defines different theories. $f(R)$ gravity reveals a class of theories defined as arbitrary function of $R$. We can consider it as a simplest example of Extended theories of gravity. Extended theories of gravity (alternative theories of gravity) evolved from the exact starting points probed by Einstein and Hilbert. Extended theories of gravity proposed to change the gravitational side of the equation \cite{2011PhR...509..167C,2008GReGr..40..357C}; instead of discovering correct calculations for the matter side of the Einstein field equations, which include inflation, DE, dark matter, large scale structure and quantum gravity.
\paragraph{}$f(R)$ gravity was first proposed by Hans Adolph Buchdahl in 1970 (although $\phi$ was used rather than $f$ for the name of arbitrary function). A wide range of phenomena can be produced from this theory by acquiring different functions \citep{Capozziello2003,Carroll2004,Chiba2003,Soussa2004,Nojiri2003}. $f(R)$ gravity provides natural gravitational alternative to DE. $f(R)$ theory of gravity helps us to describe the evolution of the universe. \cite{PhysRevD.70.043528} explained the presence of late time cosmic acceleration of the universe in $f(R)$ gravity. According to \cite{1742-6596-66-1-012005,PhysRevD.78.046006} a unification of the early time inflation and late time acceleration is permitted in $f(R)$ gravity. \cite{PhysRevD.79.083534} investigated FRW models in $f(R)$ theory of gravity. \cite{Pawar2015,Pawar2015a} have studied Abdussattar's model in $f(R,T)$ theory of gravity. \cite{Pawar2016} obtained the solutions of LRS Bianchi type $I$ cosmological model in the presence of DE with the help of $f(R,T)$ theory. \cite{Pawar2016a} have discussed axially symmetric space-time in $f(R,T)$ theory of gravity. \cite{AGRAWAL2017, 2017NewA...54...56A} have examined plane symmetric and Bianchi type $V$ cosmological models in the presence of $f(R,T)$ theory of gravity. \cite{Soumya2014} obtained a constant curvature spherically symmetric vacuum collapse in $f(R)$ theory of gravity. \cite{2010Ap&SS.330..183S} have solved the vacuum field equations for Bianchi type $I$, $III$ and Kantowski-Sachs space-times, in $f(R)$ gravity. \cite{Katore} presented Bianchi type $VI_{0}$ cosmological model to discuss the nature of the universe in $f(R)$ gravity.
\paragraph{}FRW metric represents a homogeneous and isotropic expanding or contracting universe. This metric is also called as, the standard model of modern cosmology. Also the metric is an exact solution of Einstein's field equations of GR. \cite{Chirde2012} have examined spatially homogeneous isotropic FRW cosmological model in self creation theory. \cite{Katore2012} has explored FRW cosmological model with strange quark matter in GR. In recent years study of space-time symmetries is more interesting topic. These space-time have been studied by \cite{Jain2012,Rao2009}.
\paragraph{}Motivated from the above done works, here we have studied vacuum solutions of spatial homogeneous and isotropic flat FRW metric and axially symmetric space-time in the presence of $f(R)$ gravity. We have obtained the solutions of the field equations by assuming power-law relation between scalar field $F$ and scale factor $a$. We also considered that scalar expansion ($\theta$) is proportional to the shear scalar ($\sigma$). In the last section physical behavior of both the models have been discussed with the help of physical parameters.
\section{$f(R)$ Gravity Formalism}
 As explained above $f(R)$ theory of gravity is the generalization of general relativity. The action for $f(R)$ theory of gravity is given by
\begin{equation}\label{eq:e1}
S = \frac{1}{2k^2} \int d^4 \sqrt{-g} f(R)  + \int  d^4x L_m (g_{\mu\nu}, \psi_{m} )
\end{equation}
here, $f(R)$ is a general function of Ricci scalar, $k^2 = 8 \pi G = 1$, $g$ is the determinant of the metric $g_{\mu\nu}$ and $L_m$ is the metric Lagrangian which depends on $g_{\mu\nu}$ and the field $\psi_m$. It should be noted that this action is obtained just by replacing $R$ by $f(R)$ in the standard Einstein-Hilbert action. Standard metric formalism and Palatini formalism are two formalisms which are applied to derive the field equations in $f(R)$ gravity \citep{Carroll2006,Dolgov2003a,Amendola2007c,Amendola2007b,Amendola2007a,Ruggiero2007,Allemandi2007}.\\
Now, by varying the action (\ref{eq:e1}) with respect to the metric $g_{\mu\nu}$, the corresponding field equations in $f(R)$ gravity is given as
\begin{equation}\label{eq:e2}
F(R)R_{\mu \nu} - \frac{1}{2}f(R)g_{\mu \nu} - \triangledown_{\mu} \triangledown_{\nu} F(R) + g_{\mu \nu} \square F(R) =  T^{m}_{\mu \nu} \\
\end{equation}
where 
\begin{equation}\label{eq:e3}
F(R)\equiv \frac{df(R)}{dR}, \quad \square\equiv\triangledown^{\mu}\triangledown_{\nu}
\end{equation}
$\triangledown_{\mu}$ is the covariant derivative and $T^{m}_{\mu\nu}$ is the standard minimally coupled stress energy tensor derived from the Lagrangian $L_{m}$. These are fourth order PDE's in the metric tensor $g_{\mu\nu}$, due to the last two terms on the left hand side of the equation.\\
These equations reduce to the field equations of GR if we take $f(R)$ $\equiv$ $R$. Contracting the field eqs (\ref{eq:e2}), it follows that 
\begin{equation}\label{eq:e4}
F(R)R - 2f(R) + 3\square F(R) = T
\end{equation}
In vacuum, it reduces to
\begin{equation}\label{eq:e5}
F(R)R - 2f(R) + 3\square F(R) = 0
\end{equation}
here, we considered the vacuum field equations (i.e. $T^{\mu}_{\nu}$ = 0). By rearranging eq (\ref{eq:e5}) we get.
\begin{equation}\label{eq:e6}
f(R) = \frac{F(R)R}{2} + \frac{3}{2}\square F(R)
\end{equation}
this gives an important relationship between $f(R)$ and $F(R)$ which will be used to solve the field equations and investigate $f(R)$. By putting value of $f(R)$ in the vacuum field eqs (\ref{eq:e2}) we get,
\begin{equation}\label{eq:e7}
\frac{1}{4}[F(R)R - \square F(R)]  = \frac{F(R)R_{\mu\nu} - \triangledown_{\mu}\triangledown_{\nu} F(R)}{g_{\mu\nu}}.
\end{equation}
The left hand side of eq (\ref{eq:e7}) is independent on the index $\mu$ thus the field equation can be expressed as 
\begin{equation}\label{eq:e8}
\alpha_{\mu} = \frac{F(R)R_{\mu\mu} - \triangledown_{\mu}\triangledown_{\mu} F(R)}{g_{\mu\mu}}
\end{equation}
this equation is independent of the index $\mu$ and hence $ \alpha_{\mu} - \alpha_{\nu} = 0$ for all $\mu$ and $\nu$, where $\alpha_{\mu}$ is just a notation for the traced quantity.\\
   
\section{FRW MODEL AND IT'S SOLUTIONS}
In this section, we will find exact solutions of the FRW space-time in $f(R)$ gravity. For the sake of simplicity, we take the vacuum field equations.

\subsection{FRW Space-time}
We consider the spatial homogeneous and isotropic flat Friedmann-Robertson-Walker(FRW) space-time
\begin{equation}\label{eq:e9}
 ds^{2} = -dt^{2} + A^{2}(t) dx^{2},
\end{equation}
where $A$ is a function of cosmic time $t$. \cite{Santripti} have briefly reviewed the Friedmann equations of standard model of cosmology. \cite{Naidu} have obtained a spatially homogeneous and isotropic FRW viscous fluid cosmological model in $f(R,T)$ gravity. \cite{2012EPJC...72.2203M} have studied the M37 - model and presented its action, Lagrangian and equations of motion for the FRW metric case in $f(R,T)$ theory. The corresponding Ricci scalar curvature for flat FRW model is given by
\begin{equation}\label{eq:e10}
R = 6\left[\left({\frac{\dot{A}}{A}}\right)^{2} +  \left(\frac{\ddot{A}}{A}\right) \right],
\end{equation}
where (.) represent derivative with respect to time $t$. We define the average scale factor $a$ as
\begin{equation}\label{eq:e11}
a= \sqrt[3]{A^3}\Rightarrow a=A,
\end{equation}
and the volume scale factor as
\begin{equation}\label{eq:e12}
V = a^3= {A^3}.
\end{equation}
The generalized mean Hubble parameter $H$ is defined as,
\begin{equation}\label{eq:e13}
H = \frac{1}{3}\left(H_1 +H_2 + H_3\right),
\end{equation}
where, $H_1 =H_2 =H_3 =$ $\frac{\dot{A}}{A}$ are the directional Hubble parameters in the directions of the $x$, $y$ and $z$-axes, respectively. Using eqs (\ref{eq:e11}) to (\ref{eq:e13}) we get,
\begin{equation}\label{eq:e14}
H = \frac{1}{3}\frac{\dot{V}}{V} = \frac{1}{3}\left(H_1 +H_2 + H_3\right)=\frac{\dot{a}}{a}.
\end{equation}
The expansion scalar $\theta$ is defined as
\begin{equation}\label{eq:e59}
\theta = 3H = H_{1}+H_{2}+H_{3} = \frac{3\dot{A}}{A},
\end{equation}
whereas the anisotropy parameter of the expansion $\Delta$ and shear scalar $\sigma$ are defined as
\begin{equation}\label{eq:e57}
\Delta = \frac{1}{3}\sum_{i=1}^{3}\left(\frac{H_{i} -H }{H} \right)^{2},
\end{equation}
\begin{equation}\label{eq:e58}
\sigma^{2} = \frac{1}{2}[\sum_{i=1}^{3} H_{i}^{2} -3 H^{2}] = \frac{3}{2}\Delta H^{2},
\end{equation}
where $ H_{i} $ denotes the directional Hubble parameters in the directions $x$, $y$ and $z$ axes respectively. Since eq (\ref{eq:e7}) does not depend on index $\mu$ thus, from $ \alpha_{\mu} - \alpha_{\nu} = 0$ for all $\mu$ and $\nu$ we can get the corresponding field equations in $f(R)$ theory of gravity for the metric (\ref{eq:e9}) with the help of eq (\ref{eq:e8}) as
\begin{equation}\label{eq:e15}
\left(\frac{\ddot{F}}{F} \right) + \left(\frac{\dot{A}\dot{F}}{AF}\right)+ 2\left({\frac{\dot{A}}{A}}\right)^{2}- 2 \left(\frac{\ddot{A}}{A}\right) = 0,
\end{equation}
where overhead dot (.) denotes derivative with respect to time $t$.
Eq (\ref{eq:e15}) is the non-linear differential equation with two unknowns $A$ and $F$. 

\subsection{Solution of the Field Equations}
In order to solve above equation we have used the relation between  scalar field $F$ and scale factor $a$.
\cite{Johri1994} has already been used the power-law relation between scale factor and scalar field in the context of Robertson-Walker Brans-Dicke models. However, in a recent paper \cite{0264-9381-24-15-012} have found a result in the reference of $f(R)$ gravity which shows that
\begin{equation}\label{eq:e16}
F\propto a^m,
\end{equation}
where $m$ is an arbitrary constant. Thus using power-law relation between $F$ and $a$, we have
\begin{equation}\label{eq:e17}
F = ha^m,
\end{equation}
where $h$ is the constant of proportionality and $m$ is any integer, hence it can have any integer value. For the present study, we chose the special case of $m = 3$ to simplify our calculations. Similar approach have been used by various authors for instance \cite{0264-9381-26-23-235020} have considered a case of $m = -2$ and \cite{Katore2016} considered special case of $m = -4$. In particular for $m=3$, eq (\ref{eq:e11}) and (\ref{eq:e17}) gives,
\begin{equation}\label{eq:e18}
F = hA^3,
\end{equation}
The deceleration parameter $q$ in cosmology is the measure of the cosmic acceleration of the universe expansion and is defined as
\begin{equation}\label{eq:e19}
q = - \frac{\ddot{a}a}{\dot{a}^2}.
\end{equation}
Putting value of (\ref{eq:e18}) in (\ref{eq:e15}) and after solving we get,
\begin{equation}\label{eq:e20}
A = (c_1 t + d_1)^{\frac{1}{12}},
\end{equation}
where $c_1, d_1$ are constants of integration. From eq (\ref{eq:e18}) and (\ref{eq:e20}) we get 
\begin{equation}\label{eq:e21}
F = h(c_1 t + d_1)^{\frac{1}{4}}.
\end{equation}
\subsection{Some Physical Parameters}
 In this section we will discuss the physical behavior of the model with the help of some physical parameters like Hubble parameter H, volume V, shear scalar $\sigma$, scalar expansion $\theta$, etc. as follows.\\
Using eq (\ref{eq:e20}) the mean Hubble parameter and directional Hubble parameters in the direction of $x$, $y$, and $z$-axis are given as
\begin{equation}\label{eq:e22}
H = H_1 = H_2 = H_3 = \frac{1}{12}(c_2t + d_2)^{-1}.
\end{equation}
Substituting value of $A$ from eq (\ref{eq:e20}) into (\ref{eq:e12}) we get the required volume. The volume $V$ of the universe is given by
\begin{equation}\label{eq:e23}
V = (c_1 t + d_1)^{\frac{1}{4}}.
\end{equation}
We know relation between $\theta$ and Hubble parameter $H$ as $\theta = 3H$, thus the expansion scalar $\theta$ is given by,
\begin{equation}\label{eq:e24}
\theta = \frac{1}{4}(c_2t + d_2)^{-1}.
\end{equation}
Substituting eq (\ref{eq:e20}) in eqs (\ref{eq:e57}) and (\ref{eq:e58}), we get the values of the mean anisotropy parameter $\vartriangle$ and the shear scalar $\sigma$ respectively as
\begin{equation}\label{eq:e25}
\vartriangle = 0,
\end{equation}
and
\begin{equation}\label{eq:e26}
\sigma = 0.
\end{equation}
After solving eq (\ref{eq:e19}) we get the deceleration parameter as 
\begin{equation}\label{eq:e60}
q = 11 = \rm{constant}.
\end{equation}
From eq (\ref{eq:e10}) and eq (\ref{eq:e20}) the Ricci scalar for FRW model is given by
\begin{equation}\label{eq:e27}
R = \frac{5c_1^2}{12}(c_1t + d_1)^{-2}.
\end{equation}
Substituting eq (\ref{eq:e27}) in (\ref{eq:e6}) we derived the function of Ricci scalar $f(R)$ as
\begin{equation}\label{eq:e28}
f(R) = \frac{19}{48}h(c_1)^{2}(c_1t + d_1)^{\frac{-7}{4}}.
\end{equation}
\subsection{Discussion for FRW Space-time:-}
Few important cosmological physical parameters for the solutions such as Hubble's parameter $(H)$, expansion scalar $(\theta)$, volume $(V)$, deceleration parameter $(q)$, Ricci scalar $(R)$ and function of Ricci scalar $(f(R))$ are evaluated for this model. In this case, except the deceleration parameter, all above explained parameters are function of time $(t)$. 
The expansion scalar $(\theta)$ gives the rate of expansion or contraction of the model and $\theta \rightarrow 0$ as $t \rightarrow \infty$. Hence the universe is expanding with slow rate with increase in time. However, mean Hubble's parameter $(H)$ and Hubble's parameters in the direction of $x$, $y$ and $z$ axis have finite value as $t \rightarrow 0$ and $ H\rightarrow 0 $ when $ t\rightarrow \infty $, it reveals that depending on the signature of the parameter, the rate of expansion is accelerated or decelerated. The spatial volume $(V)$ is finite at initial epoch and it increases with increase in cosmic time. The anisotropic parameter $(\triangle)$ and shear scalar sigma $(\sigma)$ are vanished. $\lim\limits_{t\rightarrow \infty} \left(\frac{\sigma}{\theta}\right)^{2}=0$, which shows that the model is isotropic throughout the evolution of the universe.

\section{AXIALLY SYMMETRIC SPACE-TIME AND IT'S SOLUTIONS}
\subsection{Axially symmetric space-time}
We consider axially symmetric space-time as,
\begin{equation}\label{eq:e29}
ds^{2} = dt^{2} - A^{2}\left[ d\chi^{2} + \alpha^{2}(\chi)d\phi^{2} \right] - B^{2}dz^{2},
\end{equation}
where $A$ and $B$ are functions of cosmic time $t$ alone and $\alpha$ is function of $\chi$. The corresponding Ricci scalar curvature for axially symmetric space-time is given by,
\begin{equation}\label{eq:e30}
R = 2 \left[- \frac{\alpha_{11}}{A^{2}\alpha} + \frac{\dot{A}^{2}}{A^{2}} + 2\frac{\ddot{A}}{A} + \frac{\ddot{B}}{B} + 2\frac{\dot{A}\dot{B}}{AB} \right].
\end{equation}
Since eq (\ref{eq:e8}) does not depend on index $\mu$. So from $ \alpha_{\mu} - \alpha_{\nu} = 0$ for all $\mu$ and $\nu$, we can get the corresponding field equations in $f(R)$ theory of gravity for the metric (\ref{eq:e29}) with the help of eq (\ref{eq:e8}) as
\begin{equation}\label{eq:e31}
\frac{\ddot{A}}{A} + \frac{\ddot{B}}{B} - \frac{\ddot{F}}{F}+ \frac{\alpha_{11}}{A^{2}\alpha} - \frac{\dot{A}\dot{B}}{AB} - \frac{\dot{A^{2}}}{A^{2}}  + \frac{\dot{A}\dot{F}}{AF} - \frac{F_{11}}{FA^{2}} = 0,
\end{equation}
\begin{equation}\label{eq:e32}
\frac{\ddot{A}}{A} + \frac{\ddot{B}}{B} - \frac{\ddot{F}}{F}+ \frac{\alpha_{11}}{A^{2}\alpha} - \frac{\dot{A}\dot{B}}{AB} - \frac{\dot{A^{2}}}{A^{2}}  + \frac{\dot{A}\dot{F}}{AF} - \frac{F_{1}\alpha_{1}}{\alpha A^{2}F} = 0,
\end{equation}
\begin{equation}\label{eq:e33}
2\frac{\ddot{A}}{A} - \frac{\ddot{F}}{F} - 2\frac{\dot{A}\dot{B}}{AB}  + \frac{\dot{B}\dot{F}}{BF} = 0,
\end{equation}
where overhead (.) represent the derivative with respect to time $t$ and suffix $1$ denote the derivative with respect to $\chi$.
\subsection{Solutions of The Field Equations}
Now on subtracting (\ref{eq:e33}) from (\ref{eq:e31}), we obtain
\begin{equation}\label{eq:e34}
\frac{-\alpha_{11}}{A^{2}\alpha} + \frac{\dot{A}^{2}}{A^{2}} + \frac{\ddot{A}}{A} + \frac{F_{11}}{FA^{2}} - \frac{\dot{A}\dot{F}}{AF} - \frac{\ddot{B}}{B} - \frac{\dot{A}\dot{B}}{AB} + \frac{\dot{B}\dot{F}}{BF} = 0.
\end{equation}
The function dependence of the metric together with (\ref{eq:e31}) and (\ref{eq:e32}) will imply 
\begin{equation}\label{eq:e35}
\frac{\alpha_{11}}{\alpha} = \xi^{2},
\end{equation}
where $\xi$ is constant. If $\xi = 0$ then $\alpha(\chi) = c_{1}\chi + c_{2}, \chi > 0 $, $c_{1}$ and $c_{2}$ are constants of integration. For the sake of simplicity take $c_{1} = 1$ and $c_{2} = 0.$
Hence we get $\alpha(\chi) = \chi $.\\ With the help of eq (\ref{eq:e35}), eq (\ref{eq:e34}) will reduce to         
\begin{equation}\label{eq:e36}
\frac{\ddot{A}}{A} - \frac{\ddot{B}}{B} + \frac{\dot{A}}{A}\left(\frac{\dot{A}}{A} - \frac{\dot{B}}{B} \right) - \frac{\dot{F}}{F}\left(\frac{\dot{A}}{A} - \frac{\dot{B}}{B} \right) + \frac{F_{11}}{FA^{2}} = 0.
\end{equation}
We will solve eq (\ref{eq:e36}) by using power law relation given in eq (\ref{eq:e17}) above. Hence we can write eq (\ref{eq:e17}) as
\begin{equation}\label{eq:e37}
F = z_{0}a^{m},
\end{equation}
where $z_{0}$ is the constant of proportionality. The scale factor $a$ is defined as 
\begin{equation}\label{eq:e38}
V = v_{0}a^{3}, 	
\end{equation}
without loss of generality take $v_{0}=1$ and $V$ is the volume. In a special case we assume $m = 3$, which implies
\begin{equation}\label{eq:e39}
F = z_{0} A^{2}B,
\end{equation}
which gives 
\begin{equation}\label{eq:e40}
\frac{\dot{F}}{F} = \frac{2\dot{A}}{A} + \frac{\dot{B}}{B}, \quad \frac{F_{11}}{F} = 0.
\end{equation}
Hence eq (\ref{eq:e36}) will imply
\begin{equation}\label{eq:e41}
\frac{\ddot{A}}{A} - \frac{\ddot{B}}{B} + \frac{\dot{A}}{A}\left(\frac{\dot{A}}{A} - \frac{\dot{B}}{B} \right) - \frac{\dot{F}}{F}\left(\frac{\dot{A}}{A} - \frac{\dot{B}}{B} \right)  = 0.
\end{equation}
Now 
\begin{equation}\label{eq:e42}
\frac{\ddot{A}}{A} = \frac{d}{dt}\left(\frac{\dot{A}}{A}\right) + \left(\frac{\dot{A}}{A}\right)^{2},
\end{equation}
and
\begin{equation}\label{eq:e43}
\frac{\ddot{B}}{B} = \frac{d}{dt}\left(\frac{\dot{B}}{B}\right) + \left(\frac{\dot{B}}{B}\right)^{2}.\end{equation}
Put the values from eq (\ref{eq:e42}) and (\ref{eq:e43}) in (\ref{eq:e41}) and solve  we get.
\begin{equation}\label{eq:e44}
\frac{A}{B} = exp(kt+k_{1}),
\end{equation}
where $k$, $k_{1}$ are constants of integration. Consider expansion scalar $\theta$ is proportional to the shear scalar $\sigma$ which yields $A = B^{n}$. Thus eq (\ref{eq:e44}) gives values of $A$ and $B$ as
\begin{equation}\label{eq:e45}
A = exp\left(\frac{n(kt+ k_{1})}{n-1}\right), \quad\quad n \neq 1,
\end{equation}
and
\begin{equation}\label{eq:e46}
B = exp\left(\frac{kt+ k_{1}}{n-1}\right),   \quad\quad\quad n \neq 1.
\end{equation}
Eqs (\ref{eq:e39}), (\ref{eq:e45}) and (\ref{eq:e46}) will gives value of $F$ as
\begin{equation}\label{eq:e47}
F = z_{0} exp\left( \frac{(2n + 1)(kt+k_{1})}{n-1}\right), \quad\quad n \neq 1.
\end{equation}
\subsection{Some Physical Parameters}
In this section we will discuss physical properties of the model. From (\ref{eq:e29}) the volume $V$ of the model is given as
\begin{equation}
V=A^{2}B,
\end{equation}
hence obtained volume $V$ is 
\begin{equation}\label{eq:e48}
V = exp \left( \frac{(2n + 1)(kt+k_{1})}{n-1}\right).
\end{equation}
The directional Hubble parameters in the direction of $\chi$, $\phi$ and $z$ are given as $H_{\chi} = H_{\phi} =\frac{\dot{A}}{A}$, $H_{z}= \frac{\dot{B}}{B}$ and obtained as
\begin{equation}\label{eq:e49}
H_{\chi} = H_{\phi} = \frac{nk}{n-1},
\end{equation}
and 
\begin{equation}\label{eq:e50}
H_{z} = \frac{k}{n-1}.
\end{equation}
Using eqs (\ref{eq:e49}) and (\ref{eq:e50}) the mean Hubble parameter of the model is 
\begin{equation}\label{eq:e51}
H = \frac{k(2n + 1)}{3(n-1)}.
\end{equation}
The scalar expansion $\theta$ of the model is
\begin{equation}\label{eq:e52}
\theta = \frac{k(2n+1)}{n-1},
\end{equation}
where as the anisotropy parameter of expansion is obtained as
\begin{equation}\label{eq:e53}
\Delta = \frac{2(n-1)^{2}}{(2n+1)^{2}}.
\end{equation}
Shear scalar $\sigma$ of the model is
\begin{equation}\label{eq:e54}
\sigma^{2} = \frac{k^{2}}{3}.
\end{equation}
Again here from (\ref{eq:e19}) the deceleration parameter is found to be
\begin{equation}\label{eq:e61}
q = -1. 
\end{equation}
Ricci scalar $R$ of the model is given as
\begin{equation}\label{eq:e55}
R = \frac{2k^{2}\left[n(3n+2)+1\right]}{(n-1)^{2}}.
\end{equation}
The function of Ricci scalar $f(R)$ is given as
\begin{equation}\label{eq:e56}
f(R) = \frac{z_{0} k^{2}(15n^{2}+14n+4)}{(n-1)^{2}}e^{\frac{(2n+1)(kt + k_{1})}{n-1}}.
\end{equation}
\subsection{Discussion for Axially Symmetric Space-time:-}
In this section we explain physical behavior of the model with the help of few physical parameters. For this model the average scale factor is found to be $a= exp\left(\frac{(2n+1)(kt+k_{1})}{3(n-1)}\right)$. As the exponential function never vanish for any value of $t$, thus obtained model is non-singular. The directional Hubble's parameters in the direction of $\chi$, $\phi$ and $z$ axis and mean Hubble's parameter $(H)$ are found to be constant, which implies that $\frac{dH}{dt}=0$. So for the larger value of Hubble's parameter, the expansion rate of the universe get increases. Shear scalar $(\sigma)$, scalar expansion $(\theta)$ and anisotropy parameter $(\triangle)$ are constant throughout the evolution of the universe. The spatial volume $(V)$ of the universe is finite when $t \rightarrow 0$ and it becomes infinite as $ t \rightarrow \infty $. The expression $\lim\limits_{t\rightarrow \infty} \left( \frac{\sigma}{\theta}\right)^{2}\neq0$, which reveals that the obtained model is anisotropic.
\section{Conclusions}
We have derived physical and geometrical properties of both the models in $f(R)$ theory of gravity. For FRW space-time we have obtained power law expansion model whereas for axially symmetric space-time we obtained exponential expansion model. 
We have deciphered solutions by using power law relation between scale factor $a$ and scalar field $F$ for both the space-time. In order to obtain the solutions of axially symmetric space-time we have assumed that expansion scalar $(\theta)$ is proportional to the shear scalar $(\sigma)$ that gives the relation between the metric coefficients $A$ and $B$ as $A=B^{n}$, where $n$ is arbitrary constant. 
Axially Symmetric model is non-singular i. e. there does not exist any physical singularity for this model. It reveals that the universe starts its expansion with finite volume. Also the negative value of the deceleration parameter indicates that the present universe is accelerating. FRW model approaches to isotropy and in this case the value of scale factor is $a = (c_1 t + d_1)^{\frac{1}{12}}$, which vanishes at $t_{s}= \frac{-d_{1}}{c_{1}}$. For this model the universe is decelerating.

\newcommand{\physrep}{{Physics Report}}
\newcommand{\apss}{{Astrophysics \& Space Science}}

\section{Acknowledgements}
We are thankful to the anonymous referee for his/her constructive comments to improve quality of the paper.
PKA acknowledges the Department of Science and Technology, New Dehli, India for providing INSPIRE fellowship.

\bibliographystyle{raa}\bibliography{mybibliography}

\end{document}